\documentclass[sigconf]{acmart}

\usepackage{booktabs} % For formal tables
\usepackage{tabularx}
\usepackage{graphicx}
\graphicspath{{figure/}}
\usepackage{subcaption}
\usepackage{amsmath,amsfonts}
\usepackage{color}
\usepackage{hyperref}
\usepackage{xspace}
\usepackage{breqn}
\usepackage{diagbox}
\usepackage{comment}

\newif\iffinal
\iffinal
 \newcommand{\raj}[1]{}
 \newcommand{\zliu}[1]{}
 \newcommand{\ian}[1]{}
 \newcommand{\revision}[1]{}
\else
 \newcommand{\raj}[1]{{\textcolor{green}{ Raj: #1 }}}

 \newcommand{\ian}[1]{{\textcolor{red}{ Ian: #1 }}}

 \newcommand{\revision}[1]{{\textcolor{red}{#1 }}}
\fi

\makeatletter
  \def\sectionautorefname{\S\@gobble}
  \def\subsectionautorefname{\S\@gobble}
  \def\subsubsectionautorefname{\S\@gobble}

\makeatother

\newcommand{\proj}{\texttt{MalleTrain}\xspace}
\newcommand{\jpa}{\texttt{JPA}\xspace}

\usepackage{tcolorbox,tikz}
\tcbuselibrary{skins}
\tcbuselibrary{breakable}
\newtcolorbox{rmkbox}[3][]
{
  breakable, 
  enhanced,
  colback  = #2!2,
  colframe= #2!8,
  boxsep=-0.5mm,
  borderline west={1.5mm}{0.05mm}{#3!15}, 
  #1,
}

\AtBeginDocument{%
  \providecommand\BibTeX{{%
    \normalfont B\kern-0.5em{\scshape i\kern-0.25em b}\kern-0.8em\TeX}}}

\copyrightyear{2024}
\acmYear{2024}
\setcopyright{rightsretained}
\acmConference[ICPE '24]{Proceedings of the 15th ACM/SPEC International Conference on Performance Engineering}{May 7--11, 2024}{London, United Kingdom}
\acmBooktitle{Proceedings of the 15th ACM/SPEC International Conference on Performance Engineering (ICPE '24), May 7--11, 2024, London, United Kingdom}\acmDOI{10.1145/3629526.3645035}
\acmISBN{979-8-4007-0444-4/24/05}

\begin{document}

\title{MalleTrain: Deep Neural Network Training on Unfillable Supercomputer Nodes}

\author{Xiaolong Ma}
\affiliation{%
  \institution{University of Nevada, Reno}
  \city{Reno}
  \state{Nevada}
  \country{USA}
}
\email{xiaolongm@nevada.unr.edu}

\author{Feng Yan}
\affiliation{%
  \institution{University of Houston}
  \city{Houston}
  \state{Texas}
  \country{USA}
}
\email{fyan5@central.uh.edu}

\author{Lei Yang}
\affiliation{%
  \institution{University of Nevada, Reno}
  \city{Reno}
  \state{Nevada}
  \country{USA}
}
\email{leiy@unr.edu}

\author{Ian Foster}
\affiliation{%
  \institution{Argonne National Laboratory}
  \city{Lemont}
  \state{Illinois}
  \country{USA}
}
\affiliation{%
  \institution{University of Chicago}
  \city{Chicago}
  \state{Illinois}
  \country{USA}
}
\email{foster@anl.gov}

\author{Michael E. Papka}
\affiliation{%
  \institution{Argonne National Laboratory}
  \city{Lemont}
  \state{Illinois}
  \country{USA}
}
\affiliation{%
  \institution{University of Illinois Chicago}
  \city{Chicago}
  \state{Illinois}
  \country{USA}
}
\email{papka@anl.gov}

\author{Zhengchun Liu}
\affiliation{%
  \institution{Argonne National Laboratory}
  \city{Lemont}
  \state{Illinois}
  \country{USA}
}
\email{zhengchun.liu@anl.gov}

\author{Rajkumar Kettimuthu}
\affiliation{%
  \institution{Argonne National Laboratory}
  \city{Lemont}
  \state{Illinois}
  \country{USA}
}
\email{kettimuthu@anl.gov}

\renewcommand{\shortauthors}{Ma et al.}

\begin{abstract}
First-come first-serve scheduling can result in substantial (up to 10\%) of transiently idle nodes on supercomputers.
Recognizing that such unfilled nodes are well-suited for deep neural network (DNN) training, due to the flexible nature of DNN training tasks, Liu et al.\ proposed that the re-scaling DNN training tasks to fit gaps in schedules be formulated as a mixed-integer linear programming (MILP) problem, and demonstrated via simulation the potential benefits of the approach.
Here, we introduce \proj{}, a system that provides the first practical implementation of this approach and that furthermore generalizes it by allowing it use even for DNN training applications for which model information is unknown before runtime.
Key to this latter innovation is the use of a lightweight online job profiling advisor (\jpa) to collect critical scalability information for DNN jobs---information that it then employs to optimize resource allocations dynamically, in real time.
We describe the \proj{} architecture and present the results of a detailed experimental evaluation on a supercomputer GPU cluster and several representative DNN training workloads, including neural architecture search and hyperparameter optimization. 
Our results not only confirm the practical feasibility of leveraging idle supercomputer nodes for DNN training but improve significantly on prior results, improving training throughput by up to 22.3\% without requiring users to provide job scalability information.
\end{abstract}



\maketitle

\section{Introduction}\label{sec:introduction}

%\subsection{Job Scheduling for Supercomputers}

Batch-scheduled high-performance computing (HPC) systems typically maintain a queue of runnable jobs, with the order in which queued jobs are run being determined by resource scheduling policies established by administrators to meet higher-level goals. 
For example, the largest supercomputers often implement policies to encourage capability computing, wherein they prioritize large jobs that cannot run elsewhere. 
Other criteria, such as job wait time and recent usage by a user or group, may also be considered when determining job priorities. But regardless of policy goals, the fact that jobs are typically given exclusive access to a fixed number of nodes while running means that nodes will be idle whenever the number of free nodes is less than the number needed to run the next job (as identified by policy).

Backfilling~\cite{mu2001utilization}, a method by which lower-priority, shorter, and/or smaller jobs are run on idle resources ahead of higher-priority jobs as long as they do not delay the start time of the higher-priority jobs, can reduce, but not eliminate, inefficiencies, which can be substantial.
%Supercomputers usually operate under policies that prioritize large applications that cannot run or run too slowly on small clusters. This is known as capability computing~\cite{allcock2017experience,ics20-alcf-logs,patel2020job}. 
%As smaller jobs can monopolize resources, large jobs can be blocked for significant periods.
%A commonly used solution is to set a minimum scale for jobs submitted to the scheduler and prioritize large jobs~\cite{allcock2017experience}.
%However, such a strategy unavoidably introduces idle nodes that cannot be backfilled (a.k.a., unfillable nodes). % \zliu{still not very clear why}
For example, in 2012, a comprehensive analysis of a 12-month workload trace of the Kraken supercomputer showed an average utilization of 94\%~\cite{you2012comprehensive}; a four-year study of the Blue Waters system revealed that monthly utilization rarely exceeded 80\%. \citet{jones2017workload}; and other studies have reported utilizations of around 90\% \cite{liu2023freetrain, patel2020job,allcock2017experience}.
These numbers can represent thousands of idle GPUs on large supercomputers.

%Although system utilization is an important goal, the primary mission of a leadership computing facility is to enable extreme-scale parallel jobs to take precedence. Thus, sacrificing some utilization to prioritize capability jobs is a norm. 
%Achieving high utilization of supercomputers has always been desired but challenging~\cite{jones1999scheduling}.
%Many shared computing clusters allow users to utilize excessive idle resources at lower cost or low priority, with the proviso that some or all may be taken away at any time~\cite{harlap2017proteus}. 
%From the job scheduling perspective, exploiting such dynamic resource availability and the often fluctuating markets for them requires agile elasticity and effective acquisition strategies.

%\subsection{Malleable Computing}
%Most conventional high-performance computing (HPC) applications are allocated a fixed amount of resources when a job starts, leading to overallocation and underutilization of computing nodes.
One approach to enhancing utilization in such environments is to 
%To enable more efficient and reliable computations in such an environment, we need to 
devise new approaches for structuring applications in \texttt{malleable forms} and for mapping these malleable applications to supercomputer resources.
A malleable computation adapts its degree of parallelism at runtime in response to external requests~\cite{feitelson1996toward}, for example by using checkpointing for semi-automated stop/restart~\cite{vadhiyar2003srs} or specialized languages and libraries~\cite{desell2007malleable,buchwald2015malleable,deelman2019evolution,acun2014parallel}.
If well managed, malleable applications can improve system utilization and scheduling efficiency and reduce average response times, compared with unmalleable jobs.
However, to realize these benefits, (a) malleable jobs need to be able to adapt dynamically to changing resource allocations and (b) job schedulers must be able to expand or shrink their resources to improve system utilization, throughput, and/or response times.
%From the perspective of resource demands, such an approach eases resource provision because malleable applications are more flexible on available resources. 
%GAIL - flexible on seems odd - do you mean the apps aare able to run on diverse resource.

% Xiaolong - The "flexible" here is to say the malleable apps could fit different numbers of nodes. For on-demand apps, they need a fixed amount of nodes. That's why we use a malleble app to fit into the dynamic resources which cannot be utilized by users.

% I didn't mean diverse resource, it should be flexible amount of resource.

%Moreover, it  provides opportunities to improve the utilization/scheduling efficiency of computing resources.
In practice, the rigid nature of both commonly used programming models like MPI and many current schedulers makes writing and running malleable applications a daunting task, which is why few malleable applications exist. %not common in practice.
%remained largely unrealized.
%Although they offer better support for malleability, 
%A further obstacle to malleability is the lack of management support for malleable jobs in existing batch systems.
%and therefore are incapable of leveraging their potential.

One intriguing source of malleable applications is deep neural network (DNN) training.
DNNs are being employed widely in scientific computing~\cite{aslan2020distributed,kleman2023full,ali2022fairdms,kates2019predicting,liu2020braggnn,liu2019deep}, and DNN training is becoming a major workload in today's supercomputers. 
%DNN training often uses data-parallel batch processing and is malleable by using more or less computing nodes.
%due to the data parallelism mechanism used to process a batch of data for gradient calculation, and 
%can be re-scaled to running on more or less nodes
%as one needs to checkpoint only the model weights and (in the case of stateful optimizers) optimizer state. 
Furthermore, deep learning frameworks such as AdaptDL~\cite{qiao2020pollux}, PyTorch TorchElastic~\cite{paszke2019pytorch}, and Elastic Horovod~\cite{sergeev2018horovod} enable scaling up and down the number of workers dynamically during training at modest cost without requiring a restart. 
%or resuming from checkpoints saved to durable storage. 
%The training speed of models in a distributed environment is subject to variation, influenced by a multitude of factors including computing power, network bandwidth, I/O capabilities, and the inherent characteristics of the models themselves, \autoref{fig:models_scalability} provides an illustrative comparison of the training speeds of various models within the same cluster. This workload is easily 
%Unfillable nodes in supercomputers can be considered 
A DNN training job is divided into many smaller tasks (mini-steps) that can be fitted into node$\times$time gaps in a supercomputer computing infrastructure.
In other words, DNN training workloads can in principle be structured as malleable computations. However, practical realization of this malleability requires the ability to 1) determine, quickly and accurately, what mini-steps should be configured for different batch queue states, and 2) assign resources and computations to run those mini-steps.

%Although some workloads, such as the deep neural network training can run at different scales due to the nature of data parallelism and can be easily re-scaled with minimum overhead, 
%\textbf{Challenge 1.} Despite that the DNN training is malleable, existing batch schedulers could not take advantage of this useful feature as they do not support rescale jobs based on the availability of computing nodes.

%\textbf{Challenge 2.} Even though existing DNN frameworks support rescale of training jobs, they do not know what is the optimal scaling to use based on the available computing resources.

%\textbf{Challenge 3.} and there is no interface to connect between the DNN frameworks and the batch schedulers

Liu et al.\ recently showed how, given knowledge of scheduler state, the task of identifying mini-steps can be formulated as a deterministic mixed-integer linear programming--based resource allocation problem~\cite{liu2023freetrain}.
However, while they showed via simulation that this ``FreeTrain'' approach could construct effective schedules for real scheduler traces, they did not address the second task just listed, by providing a practical implementation of their proposed approach.
This is a significant obstacle to the effective realization of malleable DNN training due to the need for several system components to coordinate and interact coherently: idle resource management, job progress monitoring, resource negotiation, and resource allocation. These components as well as their coordination are not readily available in today's job schedulers that were designed for unmalleable computing tasks. 

A second deficiency of the FreeTrain approach is that it requires users to provide accurate scaling information, such as measured throughput when using different numbers of nodes for DNN training jobs.
Providing this information is a substantial challenge because in many modern DNN training workflows, such as neural architecture search (NAS)~\cite{zoph2018learning, real2019regularized, liu2018darts} and hyperparameter tuning (HPO)~\cite{li2020system}, 
jobs are generated on the fly based on results produced in previous iterations by methods such as reinforcement learning~\cite{zoph2018learning} and Bayesian optimization~\cite{falkner2018bohb}.
Thus, even experienced DNN experts do not know all the model details beforehand, let alone their scalability characteristics.

In the work reported here, we propose and demonstrate solutions to the two obstacles to the practical realization of malleable DNN training just noted.
First, we present a malleable DNN training system architecture, \proj{}, which achieves the efficient coordination of the required idle resource management, job progress monitoring, resource negotiation, and resource allocation functions. 
%To this end, we propose a malleable DNN training framework, named \proj{}, to harness unfillable supercomputer nodes to meet surging demands of model training.
%Our  \proj{} system architecture design focuses on facilitating information sharing among all system components and carrying out the control flow to execute malleable scheduling decisions.
For instance, in order to make malleable scheduling decisions, the Resource Allocator must first get information about unfillable nodes from the batch scheduler (e.g., PBS~\cite{feng2007pbs} or Slurm~\cite{yoo2003slurm}), profiling information from a profiler, and current running and waiting DNN jobs from the job monitor; then, it needs to control a DNN scaling framework (e.g., Elastic Horovod) to execute the scheduling decisions. Throughout this process, it must also avoid negative impacts on jobs submitted to the main batch scheduler.

Second, we address the challenge of obtaining accurate scaling information by introducing a lightweight job profiling advisor (\jpa{}) to obtain automatically the information required for making resource management decisions.
\jpa{} runs experiments whenever a DNN training task starts, according to a schedule that minimizes associated costs by taking advantage of the fact that removing a node is faster than adding a node in distributed DNN training.
By thus obtaining accurate job information at modest cost, \jpa{} permits the MILP to make more accurate decisions, with significant benefits in practice.
%Although profiling introduces some unavoidable overhead, the long-term benefits outweigh these costs. 
We conducted extensive simulation evaluations using workloads from production supercomputer clusters, alongside experiments on a smaller cluster with synthetic logs derived from real Summit cluster logs. Our findings indicate that the more accurate information provided by \jpa{} allows \proj{} to achieve performance improvements of up to 22.3\% relative to FreeTrain.
In addition, it permits the scheduling of malleable DNN training applications, such as NAS and HPO, for which no performance information may be available.

This paper thus makes three important contributions. 
First, we propose a system architecture for running malleable DNNs on supercomputers, and implement \proj{} according to this architecture. Second, we propose a lightweight online profiler that employs an inverse-order profiling method to obtain accurate scalability information for dynamic DNN jobs. Third, we present results from both simulations with supercomputer traces and real-world executions on a cluster using synthetic traces that demonstrate the efficiency of these methods in harnessing previously idle nodes for DNN training---and thus the feasibility of using what may often be 10\% or more of previously unfillable supercomputer nodes for large-scale DNN training.

\section{Background} \label{chap:background_and_motivation}

We present background information on the methods used by cloud providers to support malleability, fragmented resources in HPC, FreeTrain, and HPC network topologies.

\subsection{Cloud-Preemptable Instances} \label{cloud_spot}

\begin{figure}
\centering
\includegraphics[width=0.45\textwidth]{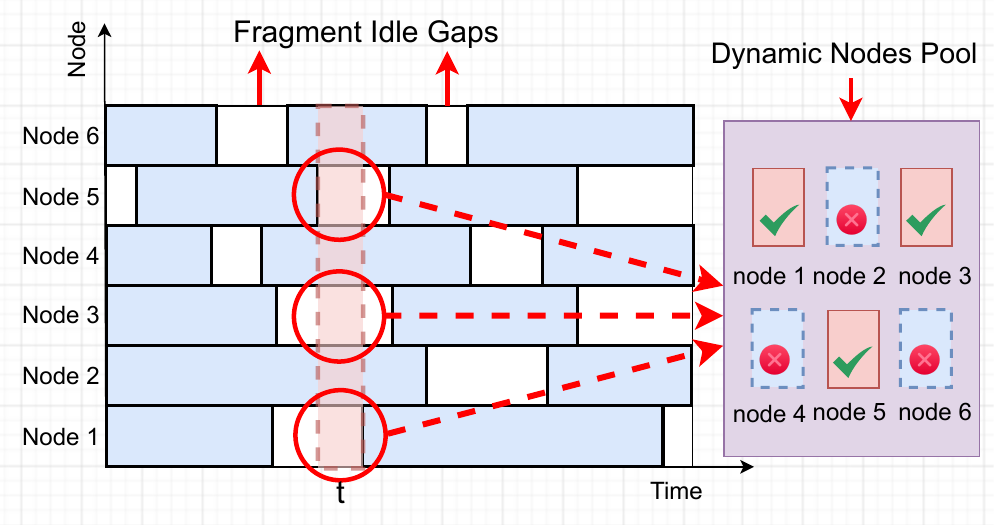}
\caption{Illustration of dynamic fragment resources on a portion of a cluster. At time $t$, there are three idle nodes in the \proj{} resource pool.}
%\ian{I am puzzling over the scenario shown here. The start times don't seem align at all, as I would expect. E.g., why doesn't the task that starts at time t+1 on node 3 start sooner? It could be because the task needs two nodes, and so it can't start until a second node becomes idle. But I don't see that. \xiaolong{can we say that there is a backfill job request 1 node just arrived? I think this could happen.} Probably it is not important, but it would be nice if there was a simple story to tell here. Another question: What is the big circle on the left of the figure meant to indicate? }\xiaolong{[3/3 Modified] figure updated, removed the circle on the left.}}
\label{fig:fragment_generation_cluster}
\end{figure}

Cloud providers such as AWS~\cite{aws}, Google~\cite{google}, and Azure~\cite{microsoft} make preemptable compute capacity available at a reduced cost via mechanisms such as AWS Spot Instances. For AWS, Spot Instances enable strategic utilization of surplus capacity; for users, they provide an opportunity to reduce their cloud expenses. To make use of such resources, however, users must be flexibile in their application runtime and tolerance for interruptions.

Spot Instances are particularly well suited for certain noncritical tasks such as data analysis, batch processing, and background operations. 
As noted, their costs are typically lower than for regular instances; on the other hand, 
%For instance, in the context of AWS Spot Instances, the pricing model diverges from that of On-Demand Instances. 
%The hourly rate of a Spot Instance is dynamic, dictated by prevailing supply and demand dynamics. This usually results in a considerably lower cost compared with On-Demand alternatives. 
they do not provide a time guarantee, introducing the possibility of unexpected interruption due to the cloud provider reclaiming running instances. To mitigate the potential impact of such interruptions, cloud providers often grant a brief time window and prior notification to clients. This advance notice enables clients to reconfigure their workload distribution, effectively rebalancing the workload across available resources. By reallocating tasks and resources in response to an impending reclamation, clients can minimize disruptions and maintain a good level of user experience. Spot Instance resources in cloud environments resemble the preemptible HPC nodes addressed by \proj{}.

\subsection{Fragment Resources on HPC}

As explored in recent research ~\cite{liu2023freetrain, patel2020job, allcock2017experience}, leadership supercomputer clusters such as Mira, Theta, and Summit exhibit utilization rates of around 90\%. Considering the substantial scale of these leadership supercomputer clusters, the unutilized resources become a significant concern. To put this in perspective, 10\% idle capacity corresponds to 460 nodes on the 4608-node Summit  and more than 1000 nodes on the 10,624-node Aurora. 
%In response, we propose harnessing these otherwise wasted resources to facilitate scalable DNN training efficiently.

Resource allocation within supercomputer clusters is typically managed by main schedulers such as Slurm~\cite{yoo2003slurm} or PBS~\cite{feng2007pbs}. These schedulers administer multiple queues to prioritize resource assignments for user requests. As depicted in \autoref{fig:fragment_generation_cluster}, inevitable fragmentary resources emerge. These fragments may not always be backfilled, and (a portion of them) may remain unassigned. However, these seemingly negligible fragments are well suited for scalable and/or fault-tolerant workloads.
%GAIL fragmentary resources of fragment resources>
The nature of these unassigned fragment resources resembles that of Spot VMs, as discussed in \autoref{cloud_spot}. In subsequent sections we will refer to these fragment resources within supercomputer clusters as \emph{preemptible nodes}. Their allocation timing lacks guarantees, rendering them unsuitable for typical fixed-size supercomputer workloads. Nonetheless, the paradigm of malleable applications, exemplified by DNN training, aligns seamlessly with this computational context. This suitability is underscored by several key factors: (1) 
%resource-intensive nature of DNN training: 
DNN training demands substantial time and computational resources; (2) 
%scalability of distributed data-parallel (DDP) training: T
the distributed data-parallel training paradigm is inherently scalable; 
(3)
%Support for Elastic Training: 
leading DNN training frameworks, such as Horovod Elastic~\cite{sergeev2018horovod} and TorchElastic~\cite{paszke2019pytorch}, adeptly support elastic training; and (4) 
%resource-Intensive Search Processes: 
DNN training often involves exhaustive searches for optimal neural network architectures and hyperparameters, consuming extensive computational resources.

The objective of \proj{} is to empower users to effectively leverage the unfilled fragments in supercomputers. Some supercomputers have a preemptable queue (the jobs submitted to this queue may be preempted anytime) explicitly to encourage the use of the unfilled nodes. A preemptable queue can be designated for \proj{} to which the users will submit adaptable DNN training jobs. \proj{} will optimally manage the allocation of unfilled nodes by dynamically expanding and shrinking these adaptable DNN jobs. To incentivize the adoption of this preemptable queue, benefits such as reduced charges,  in terms of either monetary cost or node-time consumption, can be extended to users.

\begin{table}[h]
%\vspace{-.3cm}
\centering
\caption{Queue types and their characteristics. \textit{Queue} is the queue type name on the Polaris cluster, the \textit{Min} and \textit{Max} columns give minimum/maximum number of nodes, and time, allowed per job request, and \textit{Priority} is the priority for jobs in the queue.} 
%\caption{Characteristics of resources that cannot be backfilled. \texttt{INC/h} and \texttt{DEC/h} indicate the average number of times per hour that idle nodes increased and decreased, respectively. \texttt{eq-Nodes} denotes the number of nodes that would need to be available continuously to deliver resources equal to the idle resources in \mt{}.}

\vspace{-1ex}

\begin{tabular}{c|c|c|c|r|c}
\noalign{\hrule height 2pt}
& \multicolumn{2}{c|}{\textbf{Nodes}} & \multicolumn{2}{c|}{\textbf{Time}} & \\
\textbf{Queue} & \textbf{Min} & \textbf{Max} & \textbf{Min} & \textbf{Max} & \textbf{Priority}\\\hline
debug & 1 & 2 & 5 min & 1 hr & debug \\\hline
debug-scaling & 1 & 10 & 5 min & 1 hr & debug\\\hline
demand & 1 & 56 & 5 min & 1 hr & High\\\hline
prod & 10 & 496 & 5 min & 24 hr & High\\\hline
preemptable & 1 & 10 & 5 min & 72 hr & Low\\
\noalign{\hrule height 2pt}
\end{tabular}

\label{tbl:polaris-queue-table}
\vspace{-.3cm}
\end{table}

\autoref{tbl:polaris-queue-table} displays the different queue types in the Argonne Leadership Computing Facility (ALCF) Polaris cluster~\cite{polaris}. The low job priority means that nodes allocated for the job in the preemptable queue could be reclaimed.

\subsection{FreeTrain} \label{subchap:freetrain}

As noted earlier, FreeTrain~\cite{liu2023freetrain} introduces an approach to dynamically allocating idle resources in which nodes and running job information are taken as inputs and user-defined metrics such as throughput or scalability are adopted as optimization objectives. By formulating the problem using MILP, FreeTrain is able to compute an optimal allocation of idle resources to DNN training jobs, subject to constraints such as allowed job size, feasible resource allocation, job scale information, and job migration overhead.
% TODO: Double check here, do we still need the equation for freetrain.
%\begin{align}
%\underset{x_{i,j}}{\text{Maximize}}: & \quad f(x) & \text{objective} \\
%\text{Subject to:} & \quad \sum_{x_{i,j}} & \text{job size} \\
%& \quad x - y \geq 1 & \text{resource} \\
%& \quad x - y \geq 1 & \text{job migration}
% \end{align}
%The problem scale is predetermined and theoretically attains a global optimum within a specific time. Our analysis of this method revealed that given sufficient time, the solver can yield a globally optimal solution. 

However, several practical challenges must be overcome before this approach can be realized into a production environment:

(1) \textbf{Expecting users to provide specific runtime job details can be a significant burden to users.} The MILP algorithm requires users to supply precise job-specific information, such as model training throughput and scalability, since these details serve as essential inputs for the optimization process. This requirement will significantly increase the burden on users.

(2) \textbf{Job runtimes often correlate closely with specific hardware capability and configurations.} Thus, to attain accurate job runtime information, users would have to prerun their jobs under nearly identical system settings and hardware configurations. However, this approach would be prohibitively time-consuming and resource-consuming for most supercomputer users.

(3) \textbf{In some cases, heuristic algorithms rely on current models to predict future executions, making it impractical to preprofile all potential models.} The majority of HPO/NAS algorithms are heuristic \cite{falkner2018bohb, zoph2018learning, real2019regularized, liu2018darts}, which implies that the models to be evaluated are not predetermined until the current models have completed their execution. Thus, users will not be able to provide accurate job runtime information, a situation that will lead to an invalid resource allocation plan and will largely downgrade the performance of the system.

To overcome these challenges, an intelligent online profiling mechanism is needed. Such a mechanism should accurately collect job runtime information while minimizing disruptions to the regular execution of jobs.

%Lastly, unreasonable cluster setting configurations can yield inferior solutions, significantly undermining framework performance in real-world scenarios. 

\proj{} also employs MILP to do the allocation optimization but emphasizes practical deployment aspects in supercomputer clusters. 
%\raj{not sure what does ``end-to-end workflows'' here refer to}. \xiaolong{[3/3 Modified] deleted the end-to-end workflows.} 
\jpa can be integrated seamlessly into the workflow,  orchestrating automatic profiling and obviating the need for manual input. As a result, the profiling procedure becomes an inherent facet of the process, efficiently alleviating the user from the need to provide such details beforehand. This dynamic profiling process operates in real time, eliminating the need to halt any ongoing jobs. While the profiling phase may occasionally lead to suboptimal cluster performance, we mitigate potential overhead through the implementation of a carefully designed online profiling mechanism. 
Thus our design is able to obtain accurate profiling information without  excessive operational costs.

\subsection{Topology}

\begin{figure}
\centering
\includegraphics[width=0.4\textwidth]{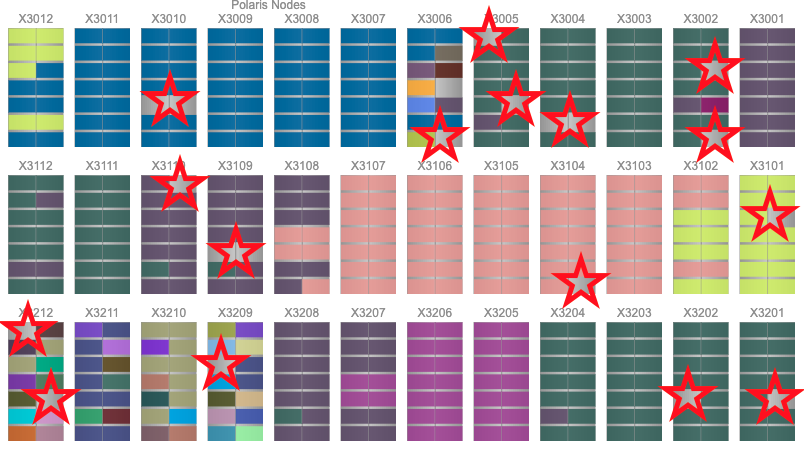}
\caption{Example of fragment resources distribution on Polaris (27th in the TOP500 supercomputer list on Nov. 2023). Red stars mark fragmented idle resources scattered on the cluster. Note: For clarity in presentation, the figure depicts a majority of the cluster rather than its entirety.}
\label{fig:topology_scatter}
\end{figure}

The network topology in a supercomputer cluster plays an important role in facilitating efficient communication, seamless data transfer, and effective management of network resources. Today, the dragonfly~\cite{kim2008technology} and fat-tree~\cite{al2008scalable} topologies are widely utilized in supercomputer clusters due to their ability to deliver high bandwidth and low latency. These features make them adept at meeting the demanding requirements of modern high-performance computing environments. The Polaris cluster and upcoming Aurora cluster in the ALCF both use the dragonfly network topology, and the Summit cluster uses fat-tree. A major concern for fragmented idle resources in a supercomputer is that such resources will often be scattered and distant from each other, as shown in \autoref{fig:topology_scatter}. 
%we take the ALCF Polaris cluster as an example, in ~\autoref{fig:topology_scatter} 
Each color represents a job; the nodes with same color were allocated to the same job. To fully utilize the inter connection bandwidth and reduce the latency, schedulers tend to assign the nodes into the same group or make them close to each other. For fragmented resources, however, usually the nodes are scattered into different topology groups. This scattering will have two major impacts. First, long distance usually means more hops are needed, which means the connections could have a higher fluctuation and cause a downgrade in the DNN training performance. Second, long distance could increase the end-to-end latency and cause more network resource contentions. We 
%address those concerns through 
perform extensive evaluation and show that the topology is not a critical bottleneck for the design of \proj{}.

\begin{figure*}[ht]
\centering
\includegraphics[width=0.7\textwidth]{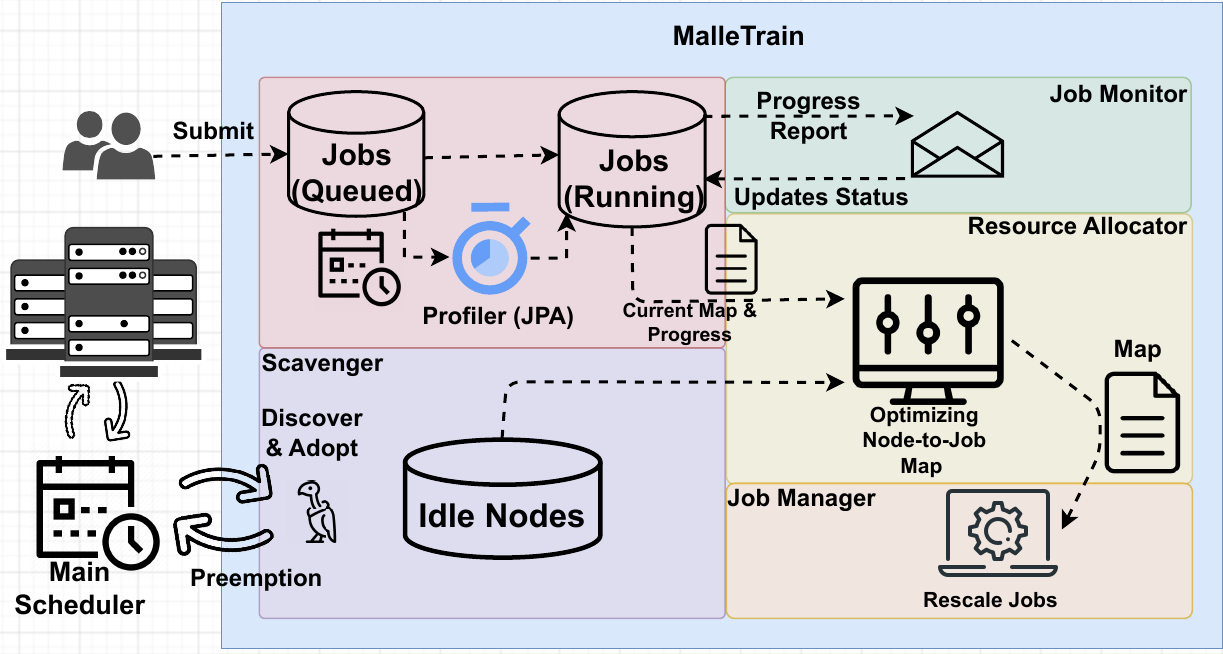}
\caption{Schematic of the \proj{} architecture. Scavenger adopts idle nodes, Resource Allocator determines a map of nodes to jobs, Job Manager rescales jobs according to the map, Job Monitor tracks job progress, and Job Profiling Advisor manages the online profiling process.}
\label{fig:sys-arch}
\end{figure*}

\section{System Design and Realization} \label{chap:system_design}

\proj{} manages the residual resources of a supercomputer cluster, in other words, those that at any particular moment have not been allocated directly by the main scheduler. Two major challenges for \proj{} arise in utilizing such residual resources: (1) their availability varies dynamically, and (2) they are preemptible. The \proj{} design enables these resources to be utilized fully for parallel DNN training. \proj{} seamlessly integrates with mainstream schedulers such as Slurm or PBS on supercomputer clusters. It operates without impacting the main scheduler, exclusively controlling the non-trivial, dynamic, residual resources that the main scheduler cannot utilize.

\subsection{System Architecture Overview}

\autoref{fig:sys-arch} shows the \proj{} architecture and its five primary components, which we describe in the following:

\vspace{1ex}
\noindent
\textbf{Scavenger} detects and collects idle nodes from the main job scheduler for \proj{}. Two primary approaches could be employed: an event-driven mechanism, whereby the main scheduler alerts \proj{} to idle nodes, or a proactive strategy, in which Scavenger periodically polls to find available unused (but preemptible when the main scheduler needs them) resources. The latter approach, preferred for its autonomy, requires no additional action from the main scheduler, ensuring seamless and efficient use of idle nodes by \proj{}.

\vspace{1ex}
\noindent\textbf{Resource Allocator} maps nodes to DNN jobs
%\bfjobs{} 
in such a way as to optimize a given metric such as throughput or scaling efficiency. 
The allocation task can be formulated as a mathematical programming problem.
In this paper we adopt the formulation of \citet{liu2023freetrain} for resource allocation.
The Resource Allocator is event-driven, with four types of events being considered: new nodes joining \proj{}, nodes being recalled by the batch scheduler (i.e., the corresponding jobs are preempted), arrival of new \proj{} jobs, and \proj{} jobs completing. 

\vspace{1ex}
\noindent\textbf{Job Manager} manages all jobs and implements the jobs-to-nodes mapping 
%the change of \ndtobfjobs{} map 
made by the Resource Allocator.

\vspace{1ex}
\noindent\textbf{Job Monitor} tracks job progress by consuming (current global batch size, timestamp) records generated by DNN training jobs via  
%To monitor the instant progress (e.g., throughput) and rescaling cost, the user needs to add 
one line of \proj{}-supplied code added to the training loop.
%to report the current global batch size and a timestamp.
The Monitor module then computes the current throughput as well as the cost incurred for each rescale operation and updates that information in a job records table to be used by the Resource Allocator.

\vspace{1ex}
\noindent\textbf{Job Profiling Advisor} manages the online profiling process, as described in \autoref{online_profiling}. The \jpa is an independent component that starts work before the job entering the Resource Allocator.

\vspace{1ex}

When nodes cannot be backfilled by the main scheduler, they are redirected to the Scavenger for utilization. Jobs submitted by users to \proj{} await the availability of nodes. As nodes become available, the jobs at the front of the queue commence execution. The running jobs transmit progress updates to the Job Monitor via a socket client. The system's architecture ensures continuous reporting of both cluster node statuses and job execution information to the Resource Allocator. The Allocator then employs MILP based on the current job distribution and number of nodes in the Scavenger. The MILP algorithm devises a strategic plan, which is represented by a map and subsequently conveyed to the Job Manager. The Job Manager then implements this plan to adjust resources accordingly. The events described in \autoref{chap:events} will trigger the Resource Allocator to run MILP and generate a new adjustment plan.

Users are provided with the option to explicitly indicate whether their job requires profiling. If so, the \jpa consults with the Resource Allocator to assess the availability of necessary node resources for profiling. Should resources be insufficient, the jobs are returned to the queue. Conversely, if adequate resources are available for profiling, the job proceeds through the profiling process. This process uniquely involves an inverse order of node numbers; further details are given in \autoref{online_profiling}. When the profiling process is done, the profiled job information will be an input to the MILP to find the optimal allocation.

\subsection{Event-Driven Resource Adjustment} \label{chap:events}

\begin{figure}[bth]
\centering
\includegraphics[width=0.4\textwidth]{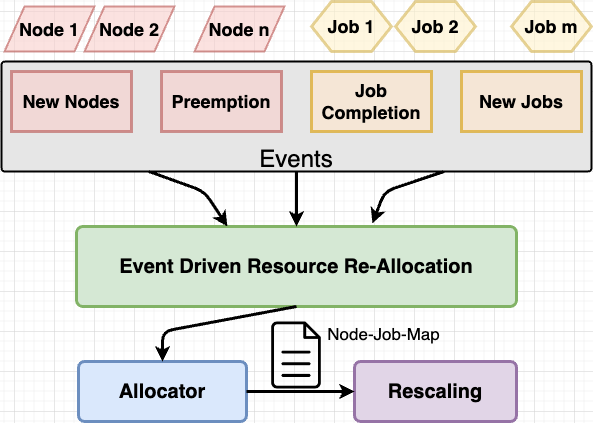}
\caption{Event-driven resource allocation process.}
\label{fig:opt-flow}
\end{figure}

Our event-driven resource management architecture is shown in \autoref{fig:opt-flow}. There are four types of events: 

\vspace{1ex}
\noindent
\textbf{New Nodes} indicates that one or more nodes have become available to \proj{}.
%added based on notifications from the main scheduler. Typically it is a batch of nodes joining \proj{} and we treat them as one event to reduce rescaling cost. 

\vspace{1ex}
\noindent
\textbf{Preemption} can be initiated at any time by the main scheduler without any prior notification. The jobs being run by \proj{} on the preempted nodes are terminated and the nodes returned to the main batch scheduler.

\vspace{1ex}
\noindent
\textbf{Job Completion}. \proj{} picks a maximum of the top (first come, first serve, FCFS) jobs from its queue to prevent excessive hunger of low-priority jobs (e.g., low-throughput jobs when sample processed per second is the target to optimize). All the selected jobs are launched by \proj{} via spawning a process using  a \texttt{subprocess} module of Python in a nonblocking fashion. The exit/completion of a job is thus notified from the Job Monitor module of \proj{}.

% TODO: here there are too many equations that correspond with the FreeTrain paper, however, these equations are not in this paper.

\vspace{1ex}
\noindent
A \textbf{New Jobs} event can trigger resource allocation only when the number of currently running jobs, $Nj_{run}$, is less than the jobs number threshold allowed in \proj{}, $Pj_{max}$. When more than one job is submitted as a batch (e.g., grid search of a hyperparameter search), $Pj_{max} - Nj_{run}$ jobs will be added to the running list as a batch to reduce the rescaling cost. When the number of arrving jobs $Nj_{arrive}$ is larger than $Pj_{max} -Nj_{run}$, the $Nj_{arrive}-(Pj_{max}-Nj_{run})$ jobs will be put into the FCFS queue for future execution.

\begin{table}[ht]
\caption{Example jobs-to-nodes map, as determined by MILP.  
Each row corresponds to a job, with scale given by the sum of the cells in the row; each column corresponds to a node, with  at most one cell in the column with value 1 indicating the job to which the node is allocated.
}
\centering
\begin{tabular}{|c|c|c|c|c|c|c|c|c|c|c|}
\hline
\multicolumn{1}{|c|}{ } & \textbf{$N_1$} & \textbf{$N_2$} & \textbf{$N_3$} & \textbf{$N_4$} & \textbf{$N_5$} & \textbf{$N_6$} & \textbf{$N_7$} & \textbf{...} & \textbf{...} & \textbf{$N_n$} \\
\hline
$J_1$ & 0 & 0 & 1 & 0 & 0 & 0 & 1 & 0 & 0 & 0 \\
\hline
$J_2$ & 0 & 0 & 0 & 0 & 0 & 0 & 0 & 0 & 1 & 1 \\
\hline
... & 1 & 0 & 0 & 1 & 1 & 0 & 0 & 0 & 0 & 0 \\
\hline
$J_4$ & 0 & 1 & 0 & 0 & 0 & 1 & 0 & 1 & 0 & 0 \\
\hline
\end{tabular}
\label{tab:node-job-map-table}
\end{table}

\vspace{1ex}

The node-job map shows the allocation plan, and ~\autoref{tab:node-job-map-table} demonstrates an example map of the allocation plan. The MILP optimizer takes the input and gives a new node-job map to the Allocator to do the reallocation. We  give  more details in \autoref{online_profiling}.

\subsection{Job Profiling Advisor} \label{online_profiling}

%\begin{figure}[t]
%\centering
%\includegraphics[width=0.45\textwidth]{figure/rescale_overhead.pdf}
%\caption{The scale up and scale down overhead on nodes with A100 GPUs}
%\label{fig:rescale-overhead}
%\end{figure}

\begin{figure}[h]
\centering
\includegraphics[width=0.45\textwidth]{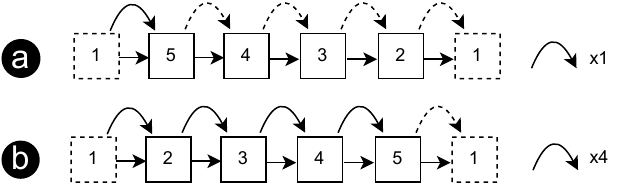}
\caption{Inverse-order rescaling sequence. The solid curve represents scale-up and the dashed curve scale-down. \jpa aims to minimize the number of scale-up operations in order to reduce overhead.}
\label{fig:rescale-seq}
\end{figure}

In contrast to traditional profiling methods that necessitate dedicated resources, our online profiling process is integrated into the training process. This approach ensures the uninterrupted operation of worker processes during profiling. The strategic design of node adjustment sequences, as depicted in \autoref{fig:rescale-seq}, to avoid scale-up operations, effectively minimizes additional overhead. Each job is equipped with a lightweight reporter (socket client), responsible for reporting job progress to the Job Monitor (socket server). This approach facilitates the automatic aggregation by the Job Monitor of the training process information that is then used for optimization purposes. Consequently, the \jpa is enabled to make precise and timely adjustments, thereby maximizing resource utilization.

\begin{figure}[th]
  \centering
  \begin{subfigure}[b]{0.8\linewidth}
    \includegraphics[width=1\columnwidth]{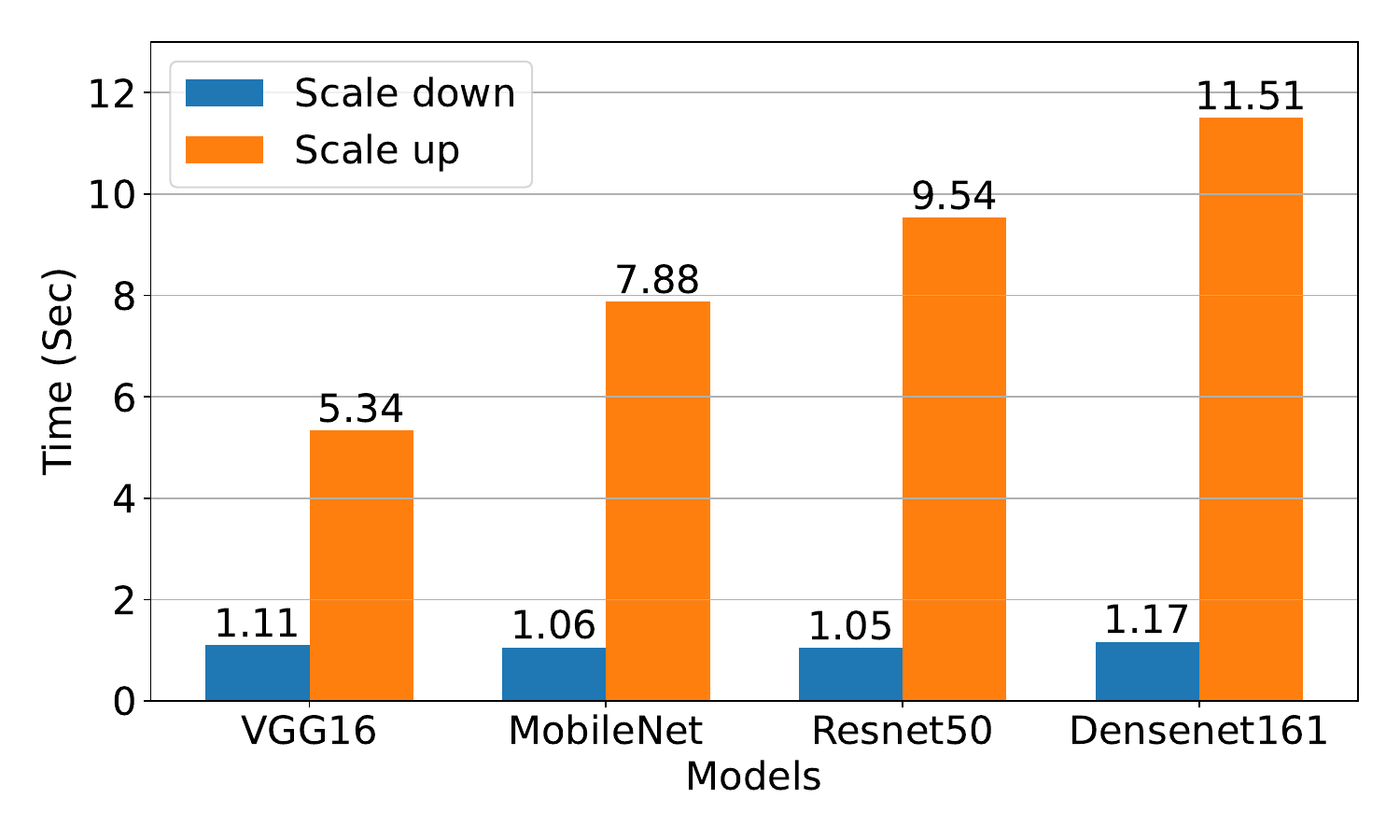}
    \caption{1-node scale-up and scale-down costs.}
    \label{fig:models_scale_up_down_comp}
  \end{subfigure}
  
  \vspace{1ex}
  
  \begin{subfigure}[b]{0.8\linewidth}
    \includegraphics[width=1\columnwidth]{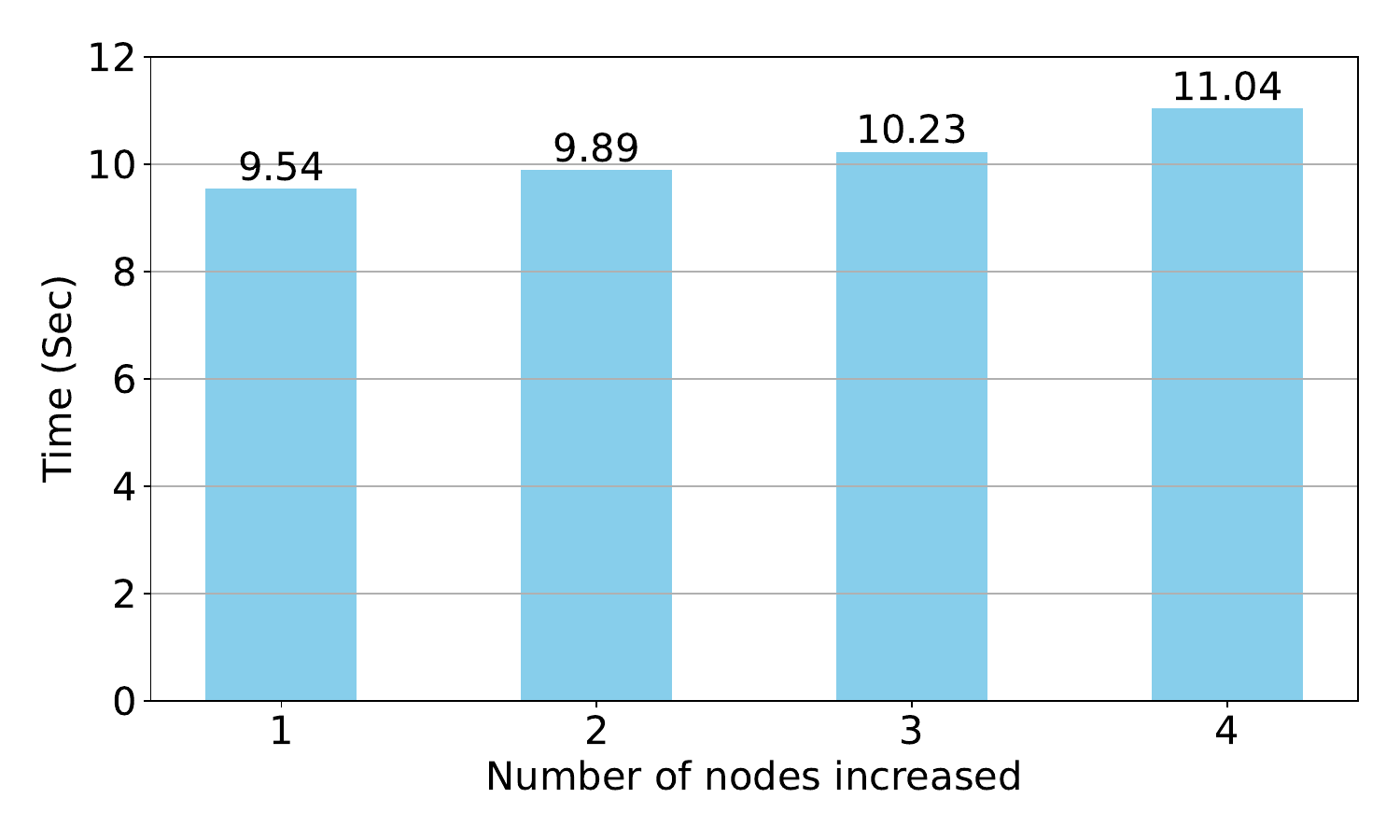}
    \caption{Scale-up times vs.\ number of nodes: ResNet-50.}
    \label{fig:increase_diff_nodes}
  \end{subfigure}
  \caption{Rescaling overhead costs on Polaris A100 GPU nodes: (a) Time to scale up and down a single node, for different models; (b) Time to scale up different numbers of nodes, for ResNet-50 model.}
  \label{fig:models_rescaling_overhead}
\end{figure}

% For users who do not obtain the models' scalability or re-scale overhead information in advance, we offer some reasonable default settings for each model, these default settings are based on experience, and although it could work, we still recommend users offer accurate input information. The inaccurate input information would downgrade the optimization results, which will adversely impact the system's overall performance. 

%To make the best use of existing resources, we placed the online profiling process in the current system. However, because of the dynamic resources required for the profile process, we need to carefully design the process to minimize disruption to ongoing jobs.

We noted in \autoref{subchap:freetrain} the necessity for online profiling in order to permit accurate MILP solutions and to handle tasks for which profile information is not available before their execution. Here we shift focus to an in-depth examination of the design elements of \jpa. In our proposed design the profiling function runs concurrently with jobs. Thus we want it to be:

\vspace{1ex}
\noindent
 \textbf{Prompt}, meaning that it processes profiling events rapidly so as to ensure efficient utilization of profiling information, and furthermore completes rapidly so as to minimize overhead and limit disruption to other tasks;

\vspace{1ex}
\noindent
\textbf{Fair}, meaning that its design incorporates principles of fairness, and that in instances where job interruption is unavoidable, a Least Recently Used (LRU) strategy is employed to ensure equitable distribution of interruptions; and

\vspace{1ex}
\noindent
 \textbf{Efficient}, meaning that it prioritizes minimal disruption to other tasks, adhering to two key principles: (1) avoiding the interruption of multiple jobs simultaneously and (2) preventing the complete cessation of any single job. %This approach has been crafted to minimize overhead and ensure process efficiency.

\begin{figure*}[ht]
\centering
\includegraphics[width=0.7\textwidth]{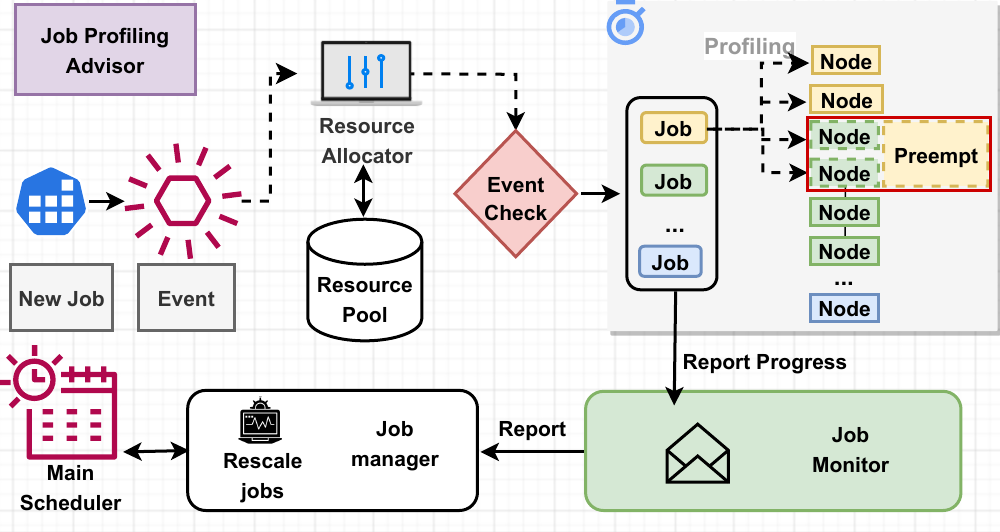}
\caption{Online profiling process.}
\label{fig:online-profile2}
\end{figure*}

Accurate MILP requires that we know, or can rapidly determine, the time that will be required to run any training mini-task on any possible number of nodes.  While obtaining this information may sound intractable, in practice the regular nature of DNN computations makes it feasible to obtain good estimates.
As in FreeTrain, we assume a fixed per-node minibatch size (when training, we employ a learning rate scheduler to adjust learning rate according to the global batch size~\cite{goyal2017accurate,you2018imagenet}).
We then need simply to measure the time per epoch for that minibatch size on different numbers of nodes, from a specified minimum to a specified maximum.

A useful optimization when performing those measurements derives from the observation that, as shown in \autoref{fig:models_scale_up_down_comp}, the cost of scaling up is consistently multiple times greater than that of scaling down. Furthermore, \autoref{fig:increase_diff_nodes} illustrates that the overhead incurred during scale-up remains relatively constant regardless of the number of nodes involved; even as the number of nodes increases, the increase in scale-up time is marginal. Consequently, in our profiling of the rescaling process, we should minimize the need for scaling up and prioritize scaling down wherever feasible. As an example, consider the two situations illustrated in \autoref{fig:rescale-seq}. If the initial number of nodes is 1 and the objective is to profile nodes 2, 3, 4, and 5, we may either: (a) scale up directly to 5 nodes and then scale down to 1, thereby gathering scalability data for all nodes using a single scale-up operation, or (b) incrementally scale up from 1 to 5, which requires four separate scale-up operations. The first approach is significantly more efficient than the first, since it requires only one scale-up operation. 
%thereby substantially outperforming the incremental method in terms of scalability profiling. 
%Since the cost of scaling up for certain models can be an order of magnitude higher than that of scaling down, the efficiency gains achieved by our method are substantial. Unlike traditional profilers that overlook the asymmetric costs associated with scaling up versus scaling down, our approach incurs significantly lower overheads, thereby enhancing its overall efficiency.

The \jpa architecture (\autoref{fig:online-profile2})  resembles that of \proj{} but with several distinctions: (1) \jpa exclusively processes new job events, since only these require profiling; in contrast, the trainer instance accepts multiple events, as described in \autoref{chap:events}. (2) The node adjustment in \jpa is decided by our profiling algorithm instead of by the MILP program. Users retain the discretion to decide whether their jobs undergo profiling. Upon receiving a profiling request from a user, a profiling event is triggered, which initiates a process whereby the Resource Allocator assesses the availability of sufficient resources for profiling. If resources are deemed adequate, a profiling job is started, temporarily preempting nodes from other jobs. Upon completion of profiling, the MILP process is engaged to make adjustments based on the newly collected information. The gathered scale information is then reported and recorded by the job manager, contributing to future optimization efforts.

%\begin{figure}
%\centering
%    \includegraphics[width=0.45\textwidth]{figure/Scalability_Fitting.png}
%\caption{Scalability fitting}
%\label{fig:scale-fit}
%\end{figure}

\begin{figure}[h]
\centering
    \includegraphics[width=0.34\textwidth,trim=5mm 6mm 5mm 13mm,clip]{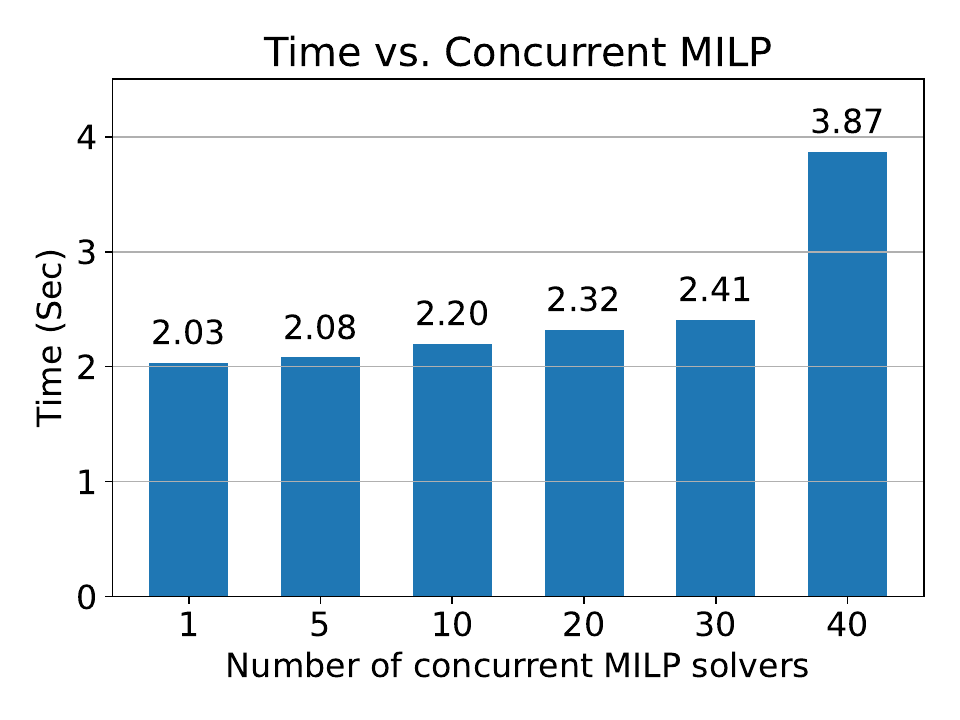}
\caption{Average time taken for an example MILP computation as the numbers of concurrent MILP computations performed on a single 32-core head node scales from 1 to 40.}
%\ian{Is my description here correct?} %\xiaolong{That's correct}}
\label{fig:num_concurrent_MILP}
\end{figure}

\subsection{Cluster Configuration}

MILP is an NP-hard problem and the cost of the MILP computation required to determine a mapping of jobs to idle nodes scales rapidly with the number of runnable jobs and available nodes.
Thus, it can be preferable to partition a supercomputer into disjoint subsets and run multiple trainers in parallel, one per subset. 
This approach restricts the maximum number of nodes to which any one job can scale, but has the advantages of reducing delays due to training and of permitting different trainers to optimize for different metrics appropriate for different task types, such as computer vision models and language models.

With multiple trainers, the question arises of whether it is advisable from a performance perspective to run more than one on a single node. 
%The scaling of jobs and nodes can significantly increase the computing time required for optimization.  Given that MILP is a well-known NP-hard problem, the overhead introduced by the frequent triggering of MILP is nontrivial, and the prolonged computation process can downgrade \proj{}'s overall performance. It is advisable to employ multiple smaller trainers in parallel rather than a single large one. Moreover, for various task types such as the training of computer vision models and language models, identifying a unified metric for optimization within a single trainer proves challenging. However, one obvious shortcoming is that the total number of available unfilled nodes has to be split among the trainers, which restricts the maximum number of nodes a job can scale to. 
%We conduct a numerical analysis to determine the optimal number of parallel trainers. This analysis will help users in configuring their number of \proj{} instances properly.
Our preliminary investigation into the effects of running multiple MILP processes concurrently on the same node revealed that the processing time begins to increase only when the number of concurrent trainers exceeds the number of cores, as illustrated in \autoref{fig:num_concurrent_MILP}. This suggests that deploying multiple trainers and running the associated MILP processes concurrently on a single head node can diminish overheads without adversely affecting the performance of standard jobs.

\section{Evaluation and Discussion} \label{chap:eval_disc}

We conducted an extensive experimental evaluation to validate the effectiveness and robustness of our framework with  real logs of supercomputer clusters. We also validated \proj{} on a small  cluster in a real production environment.

\subsection{Experiment Setup}

%\textbf{Traces} 
%In our comprehensive analysis, w
We examined trace logs from two supercomputers listed in the  TOP500 as of November 2023: Summit, ranked 7th, and Polaris, ranked 27th \cite{top500list}. The Summit log spans 14 days from February 10 to February 24, 2021, while the Polaris log covers a 7-month duration, from January 1 to July 28, 2023. 

\begin{figure}
  \centering
  \begin{subfigure}[b]{0.45\linewidth}
    \includegraphics[width=1\columnwidth]{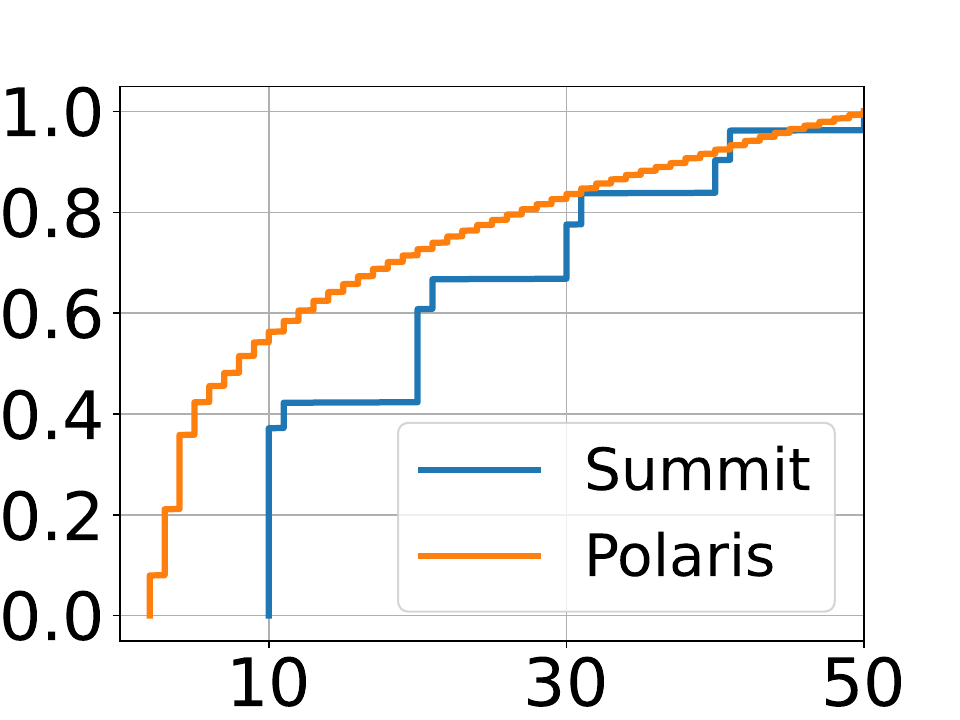}
  \end{subfigure}
  \hspace{5ex}
  \begin{subfigure}[b]{0.45\linewidth}
    \includegraphics[width=1\columnwidth]{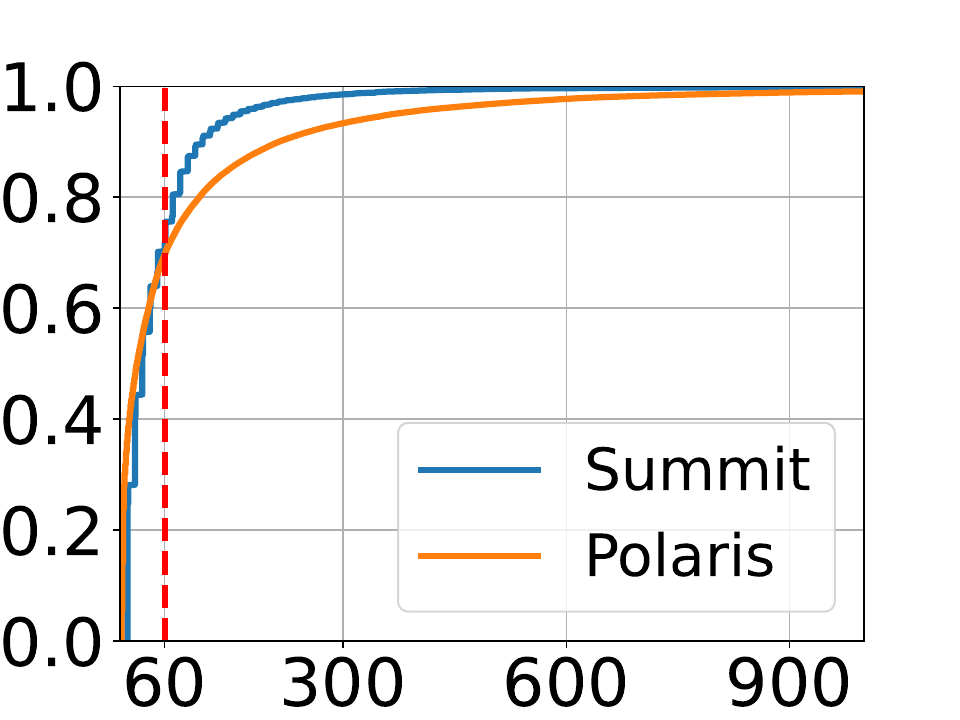}
  \end{subfigure}
  \caption{Cumulative histograms of idle gap counts on Summit and Polaris, for short gaps (0--50 secs: left) and longer gaps (0--3600 secs: right). Polaris has more shorter gaps while Summit has more gaps in the range from 60 to 600 secs.}
  \label{fig:summit_polaris_trace}
\end{figure}

\begin{figure}
  \centering
   \includegraphics[width=1\columnwidth,trim=13mm 6mm 25mm 18mm,clip]{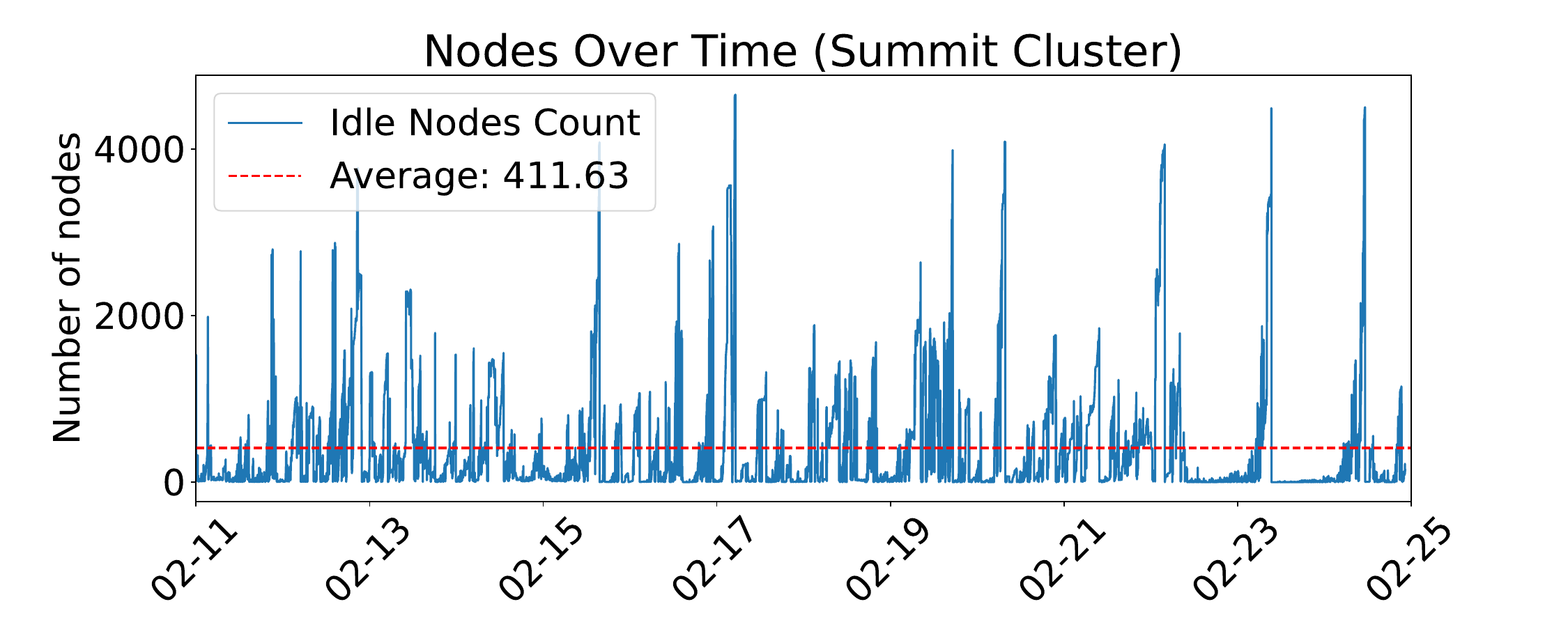}
  \caption{Idle nodes on Summit over two-week period.}
  \label{fig:summit_arrival_time}
\end{figure}

\autoref{fig:summit_polaris_trace} depicts event traces from the Summit and Polaris supercomputers. We see that Polaris has more shorter gaps than Summit, with indeed over 50\% of its event gaps being shorter than 10 seconds. A key factor contributing to this difference is Summit's policy favoring large jobs. Such jobs generally have longer durations, leading to fewer but more extended resource occupations. Conversely, without a similar policy favoring large jobs, Polaris experiences more frequent, shorter gaps between events due to the prevalence of smaller jobs. However, because of the unavailability of idle node data for the Polaris cluster, we focus on Summit trace data in our log replay simulation evaluation. \autoref{fig:summit_arrival_time}, which shows idle nodes on Summit over a two-week period, shows that the number of idle nodes varies significantly over time.

%GAIL - do you want to say anything about this two-weeek period? If not, why bother with the figure?

% Xiaolong - Hi Gail, thanks for the comments. You are right, I deleted the two weeks.

%GAIL - I don't see that the figures a and b really show much difference between Summit and Polaris. Even you say so in the text.Would the descriptioin you give be enough? 

%GAIL - you just said this The figures reveal that the Polaris cluster exhibits a higher frequency of small gap events compared with the Summit cluster. 

% Xiaolong - I replot the figures and I will put more explainations for these two figure. Here what I try to convey in (a). is that polaris has more shorter fragments (less than 10 seconds than Summit cluster). (b) shows the overall fragments distribution. 

% What I modified here. In (a) the statistic is base on fragments less than 50s and the x-axis scale changed to 0 - 50.

While plugging \proj{} into the batch scheduler of a real supercomputer would permit accurate evaluation in a real system, we would lose the ability to reproduce the same trace with different strategies, including the baseline allocation policy, for comparative research. Therefore, we instead generate representative traces and replay them on the real system for our experimental evaluation. In contrast to the simulation-based evaluation, experiments here do not rely on any performance modeling: they run the DNN training task on real supercomputer nodes. 
%GAIL - you alrady essentially said this didn't you?The only gap toward a production run is that the traces are synthesized (for reproducing with different strategies). 

A challenge for \proj{} is to optimally utilize fragmented node$\times$time resources to meet a user-specified metric (e.g., throughput in terms of samples processed per second, resource utilization/scaling efficiency).
We synthesize traces that are independent and identically distributed with real traces from supercomputers. \autoref{fig:frag-synt-pdf} compares node idle gap lengths from real Summit scheduler logs vs.\ our synthetic traces. We see that the distribution of synthetic traces is close to those of the real logs, confirming the representativeness of our synthetic traces. 

\subsubsection{Workload} 

NASBench101~\cite{ying2019bench} is a neural architecture search (NAS) benchmark dataset created to permit systematic, reproducible, and accessible evaluation of NAS algorithms. It was introduced to address the challenges associated with the high computational cost of evaluating NAS algorithms, which traditionally require training thousands of neural network architectures from scratch to find the most efficient one for a given task. 
We conducted our experiment within the search space of NASBench101, which comprises 423,624 computationally unique neural architectures. The image size for our training is 224$\times$224$\times$3. We use randomly generated tensors   instead of the real dataset to remove the potential I/O impact on our experiments. We  note that our focus here is not on the accuracy of the models but rather on assessing throughput and scalability. Varieties of deep learning models that do the HPO tasks were also evaluated in the same context as the NAS workload; the models were randomly selected from models listed in \autoref{fig:models_scalability}.

\subsubsection{Testbed} 
\begin{figure}[t]
\centering
\includegraphics[width=.4\textwidth]{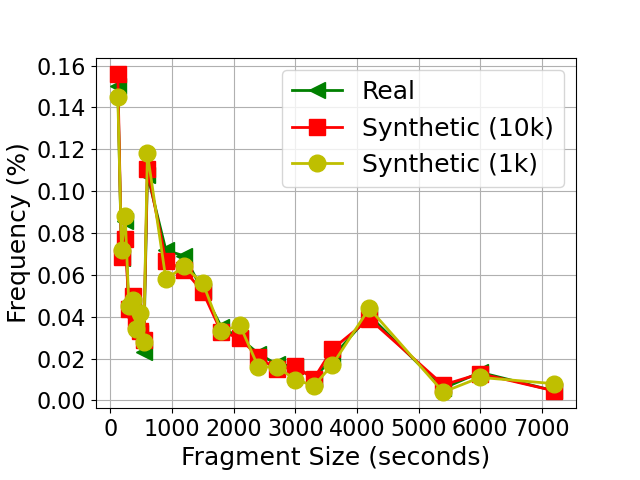}
\caption{Comparison histogram of fragment length between real logs and synthetic. Synthetic (10k) shows the statistics for 10k fragments, and Synthetic (1k) shows the statistics for 1k fragments. The  synthetic traces keep the same distribution as that of the real log.}
\label{fig:frag-synt-pdf}
\end{figure}

We conducted experiments on a 32-node cluster in which each node is equipped with four A100 GPUs. The GPUs are interconnected via NVLink within each node, and nodes are connected via InfiniBand. The synthesized traces, as depicted in \autoref{fig:frag-synt-pdf}, were instrumental in simulating the preemptive actions undertaken by the main scheduler.

\subsection{Performance Evaluation}
We conducted experiments to benchmark our system against the FreeTrain framework for preemptible resource allocation on HPC clusters. Our evaluation comprised NAS and HPO training workloads. Notably, the NAS workload exhibited more variability in training speed and scalability compared with HPO tasks.

\begin{figure}
  \centering
  \begin{subfigure}[b]{0.8\linewidth}
    \includegraphics[width=1\columnwidth]{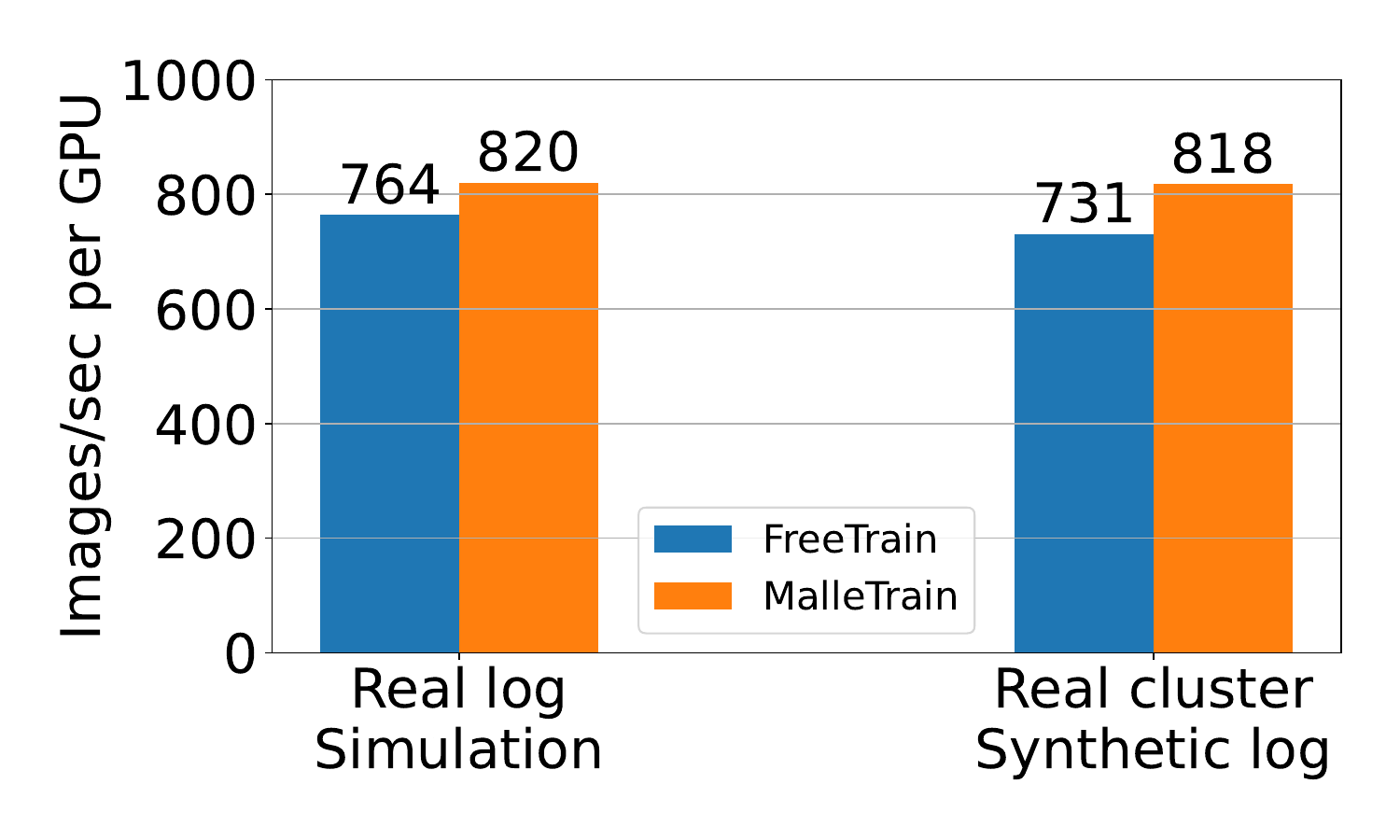}
    \caption{Neural architecture search}
  \end{subfigure}
  \begin{subfigure}[b]{0.8\linewidth}
    \includegraphics[width=1\columnwidth]{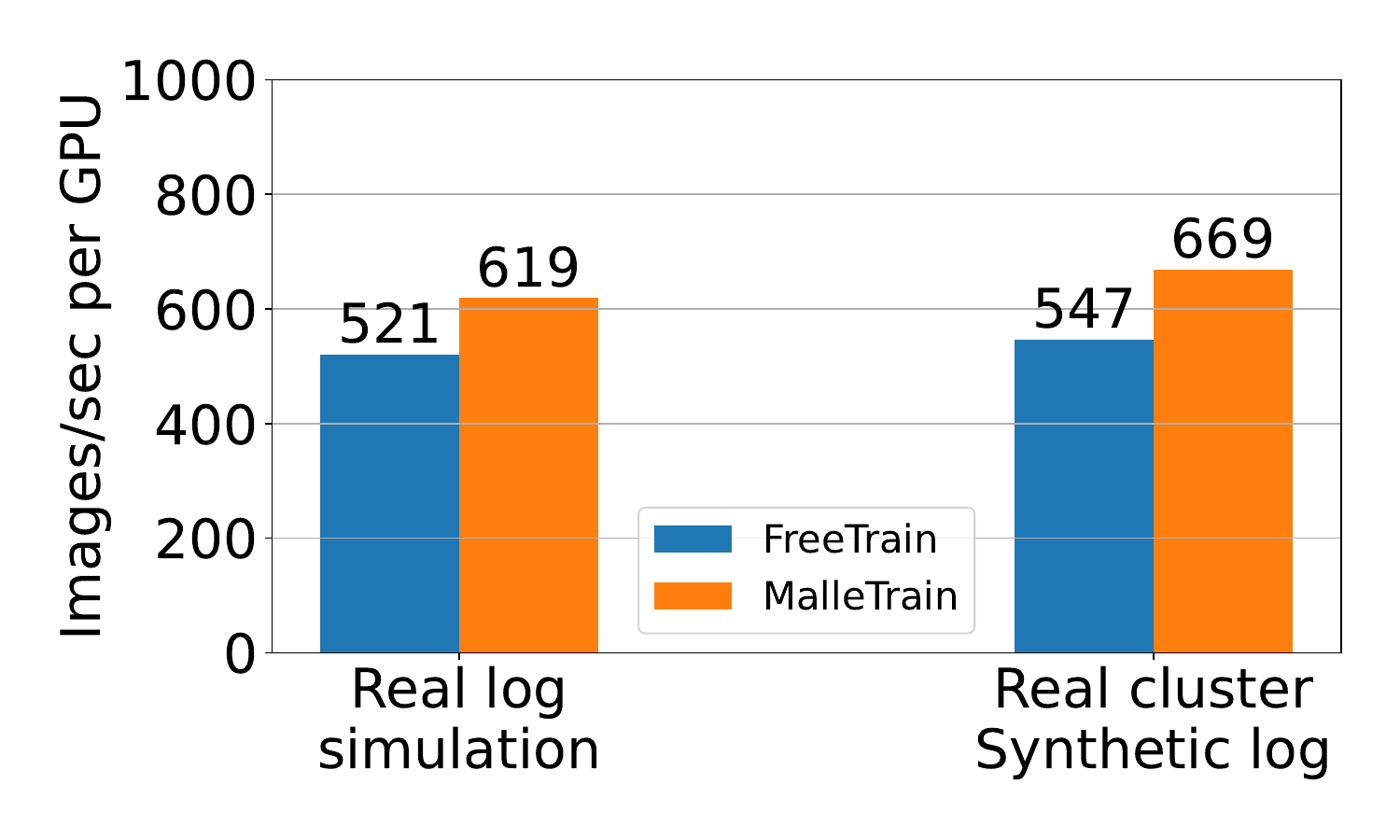}
    \caption{Hyperparameter optimization}
  \end{subfigure}
  \caption{FreeTrain vs.\ \proj{} performance for the NAS and HPO applications, as measured both with real logs on a simulator and synthetic logs on a real cluster.}
  \label{fig:freetrain_vs_malletrain}
\end{figure}

Our primary metric for comparison was the overall training throughput of the system. We ran both frameworks under identical workloads to ensure a fair comparison. For the NAS model sampling process, we randomly selected models. To maintain consistency, we set the same seed value for both frameworks, ensuring that the sequence of model training remained identical across the experiments. We conducted the simulation with the two-week log and conducted the experiments for 12 hours with the synthetic trace. The average throughput is shown in \autoref{fig:freetrain_vs_malletrain} with the NAS workload and HPO workload. We see that \proj{} outperforms FreeTrain in various settings.

\begin{figure}
  \centering
  \begin{subfigure}[b]{0.45\linewidth}
    \includegraphics[width=1\columnwidth]{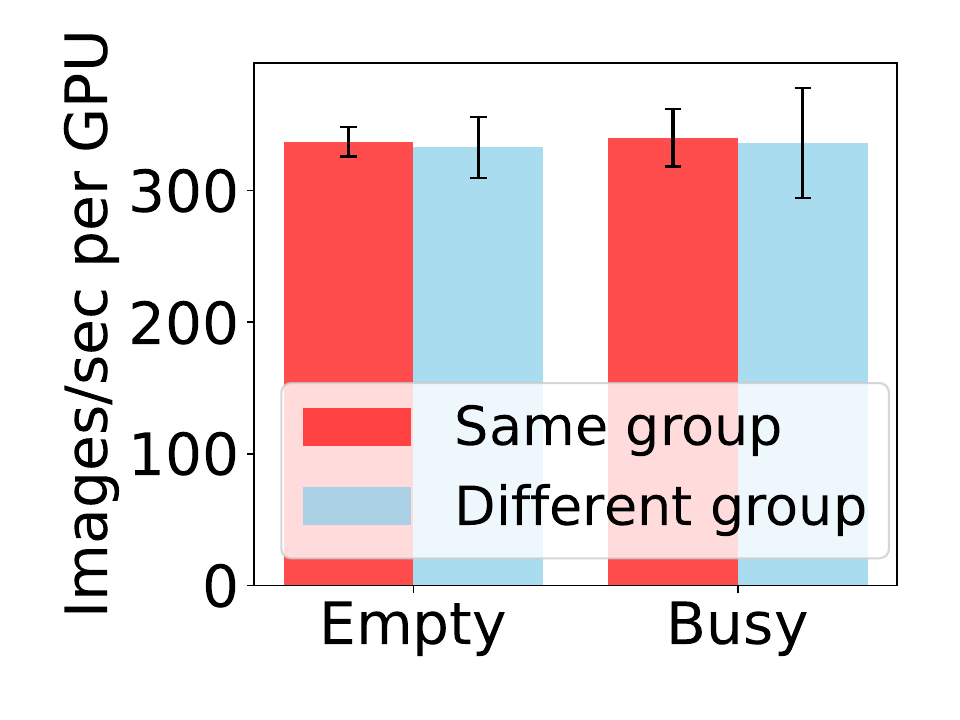}
    \caption{ResNet-50}
  \end{subfigure}
  \hspace{5ex}
  \begin{subfigure}[b]{0.45\linewidth}
    \includegraphics[width=1\columnwidth]{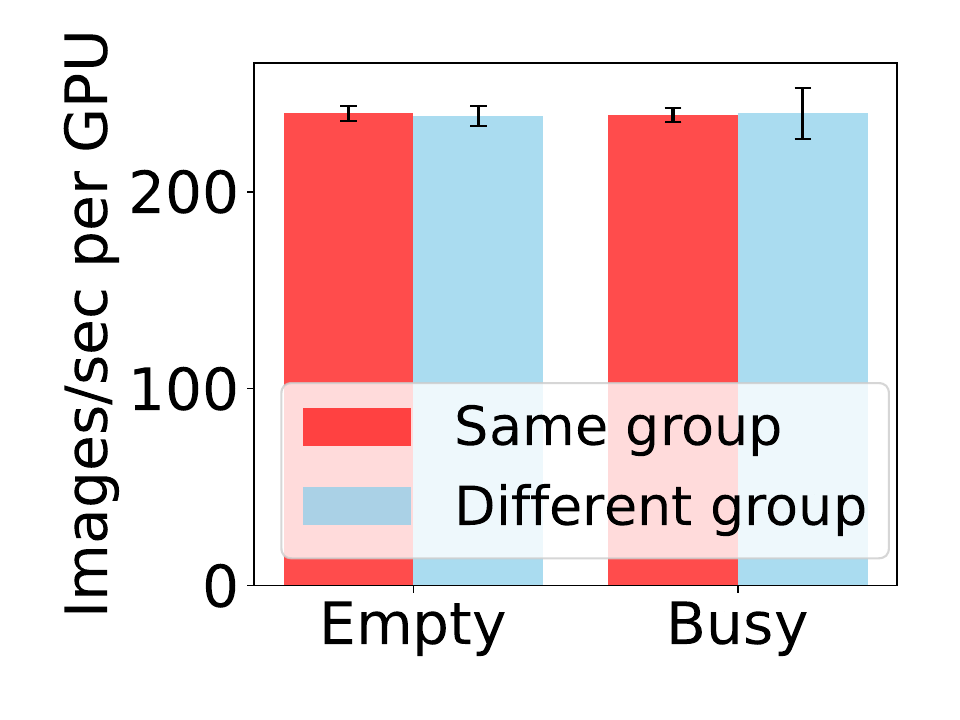}
    \caption{VGG-19}
  \end{subfigure}
  \caption{We analyzed training performance for sample \proj{} jobs under four different scenarios: \textit{Same Group, Empty} (where all nodes are located within the same Dragonfly group and the cabinet is empty), \textit{Same Group, Busy} (where all nodes are within the same Dragonfly group but are collocated with other jobs), \textit{Different Group, Empty} (where nodes are distributed across two Dragonfly groups with the two cabinets empty), and \textit{Different Group, Busy} (where nodes are distributed across two Dragonfly groups and collocated with other jobs). The results demonstrate consistent training speeds for both models across all scenarios. The error bars for the \textit{Different Group, Busy} scenario reveal higher variances in training speed, indicating fluctuations occur primarily in this scenario. However, the average training speed remains consistent despite these fluctuations.}
  \label{fig:resnet_vgg_topo}
\end{figure}

\subsection{Topology Impact Analysis}

\begin{figure}
\centering
\includegraphics[width=0.45\textwidth]{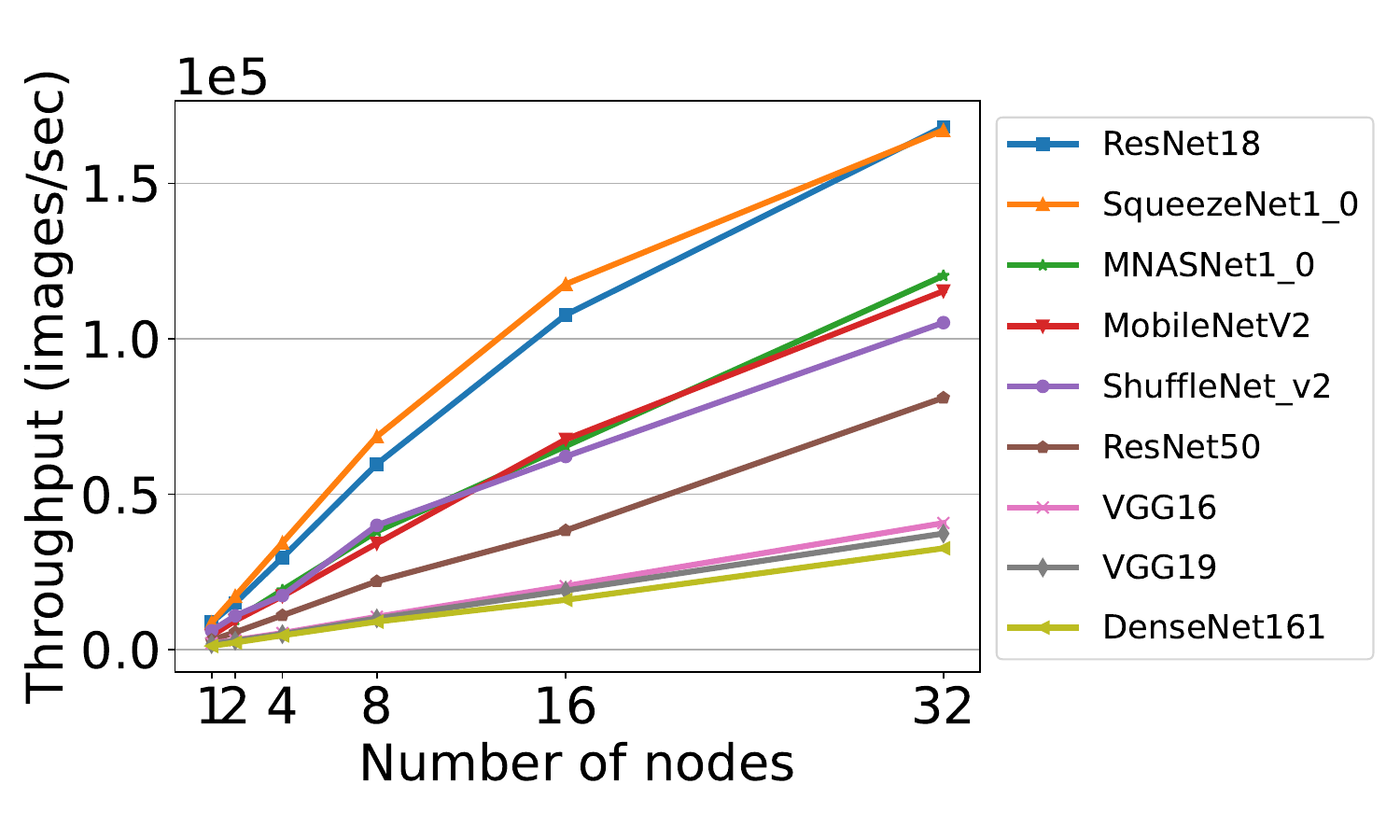}
\caption{Trend analysis of model scalability on 32 Polaris nodes, each with 4 A100 GPUs.}
\label{fig:models_scalability}
\end{figure}

The dynamic and randomly scattered nature of fragmented resources across the cluster raises concerns about potential declines in the overall performance of training jobs. To address these concerns, we conducted experiments on the Polaris cluster with the dragonfly network topology. Our study involved comparing the performance of nodes confined within a single dragonfly group versus those distributed across multiple dragonfly groups. \autoref{fig:resnet_vgg_topo} shows that the physical distribution of nodes, whether scattered or closely situated, has minimal impact on NAS/HPO DNN training speed. \autoref{fig:models_scalability} indicates robust scalability of models even at the 32-node level, each node equipped with 4 NVIDIA A100 GPUs, encompassing a total of 128 A100 GPUs.

The underlying reasons for these observations are multifaceted. First, leadership-class supercomputer clusters are typically outfitted with high-performance network devices. For instance, Polaris is equipped with the HPE Slingshot 11 interconnect, offering up to 200 Gb/s point-to-point bandwidth. Second, the networking infrastructure in these clusters is often highly overprovisioned, mitigating network contention among applications running on different nodes. Third, modern distributed deep learning frameworks, such as PyTorch~\cite{paszke2019pytorch} and Horovod~\cite{sergeev2018horovod}, effectively overlap computing and communication tasks. This overlapping functionality reduces the network's impact on training speed, thereby diminishing the sensitivity to network conditions.

% Leave related work blank for now.
\section{Related Work} \label{chap:related_work}

We have already referred to the pioneering work of Liu et al.\ on FreeTrain~\cite{liu2023freetrain}, while noting also that certain assumptions and strict requirements make it fall short in the real production environment. FreeTrain heavily relies on users to provide accurate runtime information from training jobs,  which increases the burden to the users, making it impractical for use. Indeed, in some widely used heuristic NAS/HPO algorithms, FreeTrain has to guess a configuration or provide information solely based on user experience; the inaccurate or out-of-date information might largely downgrade the overall performance of the MILP optimization algorithm. In contrast, \proj{} integrates automatic profiling components into the process and doing the profiling automatically.

Pollux~\cite{qiao2021pollux} is a resource-adaptive DNN training and scheduling framework designed to efficiently rearrange distributed deep learning processes, particularly in dynamic-resource environments such as shared clusters and cloud infrastructures. This framework employs Kubernetes for efficient scheduling, rescaling, and reconfiguring of job batch sizes and learning rates, thus maximizing training performance and optimizing resource utilization. Pollux operates on a fixed-size cluster, however, whereas \proj{} can handle dynamically varying cluster sizes.

\section{Conclusion}\label{chap:conclusion_future}
We have introduced \proj{}, a system that we demonstrate can employ idle fragmented nodes on batch-scheduled HPC systems for large-scale DNN training.
%that are otherwise wasted under existing HPC scheduler configurations .
\proj{} defines a workable architecture for efficient use of such idle nodes, and via its job profiling advisor, which efficiently gathers accurate job execution data at runtime with minimal interference to ongoing tasks, enables idle nodes to be employed efficiently even for dynamic workloads such as neural architecture search and hyperparameter optimization.
Detailed performance studies involving both simulations and experiments validate the effectiveness of the approach and show that \proj{} achieves >20\% more training throughput than was reported, on the basis of simulation studies alone, for a precursor system.
\proj{} thus opens up the feasibility of both improving the utilization of large HPC systems and increasing the resources delivered to DNN applications.
%In contrast to FreeTrain, which primarily concentrates on refining optimization algorithms, \proj{} places greater emphasis on architectural design and addressing pragmatic challenges in production environments. 
%Through efficient system design, MalleTrain achieves >20\% more training throughput than does FreeTrain while eliminating the burden on users to provide the job scalability information.
Moreover, the methodologies developed in this study have potential applications beyond their current scope. They could be adapted, for example,  to infrastructure management tasks, such as scheduling in Kubernetes clusters and other cloud computing platforms.

\section*{Acknowledgment}
We are grateful to the reviewers for their valuable feedback and comments and Gail Pieper for helping us edit our paper. This work was sponsored in part by NSF grant CAREER-2048044 and by the U.S.\ Department of Energy, Office of Science, under contract DE-AC02-06CH11357. This research used resources of the Argonne Leadership Computing Facility, a DOE Office of Science User Facility supported under Contract DE-AC02-06CH11357.

%\section{Acknowledgment}

%This work is supported in part by the following grants: DOE xxx-1756013, IIS-1838024 (using resources provided by Argonne Leadership Computing Facility as part of the DOE FreeTrain program). We thank the anonymous reviewers for their insightful feedback.


\bibliographystyle{ACM-Reference-Format}
\balance
\bibliography{BFTrainer}

\end{document}